\documentclass[12pt]{article}

\usepackage{amsmath}
\usepackage{amssymb}
\usepackage{graphicx}

\def \be{\begin{equation}}
\def \ee{\end{equation}}
\def \beq{\begin{eqnarray}}
\def \eeq{\end{eqnarray}}
\input{tcilatex}
\begin{document}

\date{\today }
\date{\today }

\bigskip

\bigskip \ 

\bigskip \ 

\begin{center}
\textbf{Hyperbolic trajectories around Black Holes}\bigskip

\smallskip

E. A. Le\'{o}n$^{a}$\footnote{%
ealeon@uas.edu.mx}, J. A. Nieto$^{a}$\footnote{%
niet@uas.edu.mx, janieto1@asu.edu}, E. R\'{\i}os-L\'{o}pez$^{b}$\footnote{%
riloemm@inaoep.mx}

\smallskip

$^{a}$\textit{Facultad de Ciencias F\'{\i}sico-Matem\'{a}ticas de la
Universidad Aut\'{o}noma}

\textit{de Sinaloa, 80010, Culiac\'{a}n Sinaloa, M\'{e}xico}

\smallskip

$^{b}$\textit{Instituto Nacional de Astrof\'{\i}sica, \'{O}ptica y }

\textit{Electr\'{o}nica (INAOE), 72840, Tonantzintla, Puebla, M\'{e}xico}

\bigskip \ 

\bigskip \ 

\textbf{\bigskip \ }

\textbf{Abstract}
\end{center}

\noindent We analyse test particle's trajectories around the geometry of a
Schwarzschild black hole. In order to resemble sections of jets in the
neighborhood of a black hole, we consider the conserved quantities
corresponding to constraints imposed on the trajectories of the test
particles, namely conic and hyperboloidic trajectories. As expected, the
energy and angular momentum are closely related to the solutions in the
non-constrained case.

\bigskip \bigskip

\ 

\bigskip \ 

Keywords: Schwarzschild geometry, Trajectories around Black holes,
Constrained systems.

Pacs numbers: 04.20.Fy, 04.20.Jb, 04.70.Bw

January 2017

\newpage

\noindent \textbf{1. Introduction}\smallskip

\noindent The orbits and trajectories of objects around black holes have
been extensively studied in the literature \cite{Frolov}-\cite{Hod}.
However, there remain several interesting scenarios where the theory can be
tested and even those where one can search for modifications of general
relativity \cite{Clifton}\cite{Capozziello}. Eventually, this could lead to
a better understanding of dark energy and dark matter \cite{Peebles}-\cite{Kleidis}.

In this work we consider hyperbolic trajectories of test particles around a
Schwarzschild black hole. We start by reviewing the formalism used for
obtaining conserved quantities in this static geometry, mentioning how the
static limit can be viewed as a hamiltonian constrained system. Next, we
turn the attention to the case of constrained trajectories of test
particles. The motivation for this analysis is that the case of a constraint
in the form of hyperboloid can be contrasted with the case of particles in
the surface of a jet in the boundaries of a black hole, at least to some
approximation.

\bigskip

\noindent \textbf{2. Static geometry for black holes as hamiltonian system}%
\smallskip

\noindent We start by considering the static geometry described by the
Schwarzschild metric, given by the metric%
\begin{equation}
ds^{2}=-\gamma dt^{2}+\gamma ^{-1}dr^{2}+r^{2}(d\theta ^{2}+\sin ^{2}\theta
d\phi ^{2}),  \tag{1}
\end{equation}%
where

\begin{equation}
\gamma =1-\frac{R_{S}}{r}.  \tag{2}
\end{equation}%
Here, the quantity $R_{S}=\frac{2GM}{c^{2}}$ is the Schwarzschild radius.
The notation in this article is as follows: greek indices ($\mu $, $\nu $,
...) run from $0$ to $3$ and latin indices ($i$, $j$, ...) run from $1$ to $%
3 $, corresponding to purely spatial indices. Derivatives respect to
coordinate $r$ will be denoted by primes, e. g. $\frac{df}{dr}=f^{\prime
}(r) $, while derivatives respect to $\tau $, the proper time, shall be
written by an over dot, as $\dot{t}=\frac{dt}{d\tau }$. Further, from here
on we consider units such that $G=c=1$.

For obtaining the trajectories one can vary the function

\begin{equation}
\mathcal{L} =\frac{1}{2}m\left[ -\gamma \dot{t}^{2}+\frac{\dot{r}^{2}}{%
\gamma }+r^{2}\left( \dot{\theta}^{2}+\sin ^{2}\theta \dot{\phi}^{2}\right) %
\right] ,  \tag{3}
\end{equation}%
where $m$ denotes the test particle mass.

Let us check the simplest case, where we impose a particle to be moving in the
plane $\theta =\frac{\pi }{2}$. Then (3) transforms into

\begin{equation}
\mathcal{L} =\frac{1}{2}m\left( -\gamma \dot{t}^{2}+\frac{\dot{r}^{2}}{%
\gamma }+r^{2}\dot{\phi}^{2}\right) .  \tag{4}
\end{equation}%
The Euler-Lagrange equation for $t$ is

\begin{equation}
\frac{d}{d\tau }(m\gamma \dot{t})=0.  \tag{5}
\end{equation}%
For the coordinate $r$, the variation leads to

\begin{equation}
\frac{\ddot{r}}{\gamma }-\frac{\dot{r}\dot{\gamma}}{\gamma ^{2}}+\frac{%
\gamma ^{\prime }\dot{t}^{2}}{2}+\frac{\dot{r}^{2}\gamma ^{\prime }}{2\gamma
^{2}}-r\dot{\phi}^{2}=0.  \tag{6}
\end{equation}%
Since we have that $\dot{\gamma}=\frac{d\gamma }{d\tau }=\frac{d\gamma }{dr}%
\frac{dr}{d\tau }=\gamma ^{\prime }\dot{r}$, we can substitute this in (6)
and multiply by $\gamma $, obtaining the simpler relation

\begin{equation}
\ddot{r}-\frac{\dot{r}^{2}\gamma ^{\prime }}{2\gamma }+\frac{\gamma \gamma
^{\prime }\dot{t}^{2}}{2}-\gamma r\dot{\phi}^{2}=0.  \tag{7}
\end{equation}

In a similar way, the variation respect to $\phi $ yields

\begin{equation}
\frac{d}{d\tau }(mr^{2}\dot{\phi})=0.  \tag{8}
\end{equation}%
The solutions for (5) and (8) are 
\begin{equation}
\dot{t}=\frac{a}{\gamma }  \tag{9}
\end{equation}%
and%
\begin{equation}
\dot{\phi}=\frac{h}{r^{2}},  \tag{10}
\end{equation}%
respectively, where $a$ and $h$ are constants. Substitution of both
solutions into (7) gives

\begin{equation}
\ddot{r}-\frac{\dot{r}^{2}\gamma ^{\prime }}{2\gamma }+\frac{a^{2}\gamma
^{\prime }}{2\gamma }-\frac{h^{2}\gamma }{r^{3}}=0.  \tag{11}
\end{equation}

Now, in this case ($\theta =\pi /2$) the Schwarzschild metric (1) can be
expressed as $-\gamma \dot{t}^{2}+\frac{\dot{r}^{2}}{\gamma }+r^{2}\dot{\phi}%
^{2}=-1$, and taking into account (9) and (10) this is the same as

\begin{equation}
\dot{r}^{2}=-\gamma +a^{2}-\frac{h^{2}\gamma }{r^{2}}.  \tag{12}
\end{equation}

The combination of (11) with (12) allow us to cancel terms with $a^{2}$, and
therefore we arrive to the relation

\begin{equation}
\ddot{r}+\frac{\gamma ^{\prime }}{2}+\frac{h^{2}\gamma ^{\prime }}{2r^{2}}-%
\frac{h^{2}\gamma }{r^{3}}=0.  \tag{13}
\end{equation}

Since $\gamma =1-\frac{2M}{r}$, we have that $\gamma ^{\prime }=\frac{2M}{%
r^{2}}$. With this, and rearranging some terms in (13), we see that it is
equivalent to

\begin{equation}
\ddot{r}+\frac{M}{r^{2}}=\frac{h^{2}}{r^{3}}\left( 1-\frac{3M}{r}\right) . 
\tag{14}
\end{equation}

\begin{figure}[h]
\begin{center}
\includegraphics[width=0.8\columnwidth]{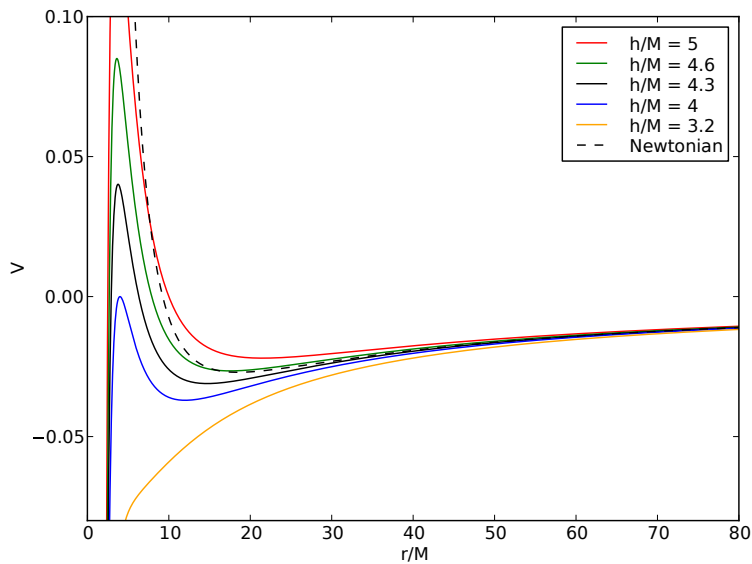}
\caption[potential]{\footnotesize The relativistic and Newtonian potentials for radial motion for different values of $h/M$.}
\label{potential}
\end{center}
\end{figure}

Multiplying (14) by $\dot{r}$ and integrating, we obtain

\begin{equation}
\frac{\dot{r}^{2}}{2}-\frac{M}{r}+\frac{h^{2}}{2r^{2}}-\frac{Mh^{2}}{r^{3}}%
=E,  \tag{15}
\end{equation}%
where $E$ is a constant, that can identified with the energy of the system
(per mass unit), and which is related with $a$ by $E=\frac{a^{2}-1}{2}$.
This allows to interpret (15) in the form $E=\frac{\dot{r}^{2}}{2}+V$, where 
$V$ is the potential $V=-\frac{M}{r}+\frac{h^{2}}{2r^{2}}-\frac{Mh^{2}}{r^{3}%
}$. In Fig. \ref{potential} is plotted the potential $V(r)$ for radial motion for different values of $h$ and also the Newtonian potential (dashed line), where can be seen that for large values of $r$ the relativistic potential is similar to the Newtonian potential. This function has extrema in $r_{\pm }=\frac{h}{2M}\left( h\pm \sqrt{%
h^{2}-12M^{2}}\right) $. If $h^{2}>12M^{2}$, this value indicates an (stable)
outer circular orbit, as well as an (unstable) inner circular orbit for test
particles.

\noindent Furthermore, the equations can be recasted to a suitable form by mean of the
change of variable $u=\frac{1}{r}$. This implies $\frac{du}{d\phi }=-\frac{1%
}{r^{2}}\frac{dr}{d\phi }$, and we also use the relation $\dot{r}=\dot{\phi}%
\frac{dr}{d\phi }=\frac{h}{r^{2}}\frac{dr}{d\phi }$, where in the last
equality we have taken into account (10). By substituting all these in (15),
we have:

\begin{equation}
\frac{h^{2}}{2}\left( \frac{du}{d\phi }\right) ^{2}-Mu+\frac{h^{2}u^{2}}{2}%
-Mh^{2}u^{3}=E.  \tag{16}
\end{equation}

Differentiating respect to $\phi $ and rearranging terms, we see that (16)
yields the simplified version

\begin{equation}
\frac{d^{2}u}{d\phi ^{2}}+u=3Mu^{2}+\frac{M}{h^{2}}.  \tag{17}
\end{equation}

Equations (10) and (17) determine the trajectories of tests particles moving
in the gravitational field imposed by Schwarzschild metric. Of course,
another approach is to solve directly the geodesic equation $\frac{%
d^{2}x^{\mu }}{d\tau ^{2}}+\Gamma _{\alpha \beta }^{\mu }\frac{dx^{\alpha }}{%
d\tau }\frac{dx^{\beta }}{d\tau }$. We can compare the mentioned equations
with the corresponding relations in newtonian theory, where we have $\frac{%
d\phi }{dt}=\frac{h}{r^{2}}$ and $\frac{d^{2}u}{d\phi ^{2}}+u=\frac{M}{h^{2}}
$, respectively. The well-known relativistic correction is the term $3Mu^{2}$
in (17), that permits to calculate the perihelion shift in Mercury, for instance \cite{Carroll}%
.\bigskip

\noindent \textbf{3. Hyperbolic trajectories.}\smallskip

\begin{figure}[h]
\begin{center}
\includegraphics[width=0.7\columnwidth]{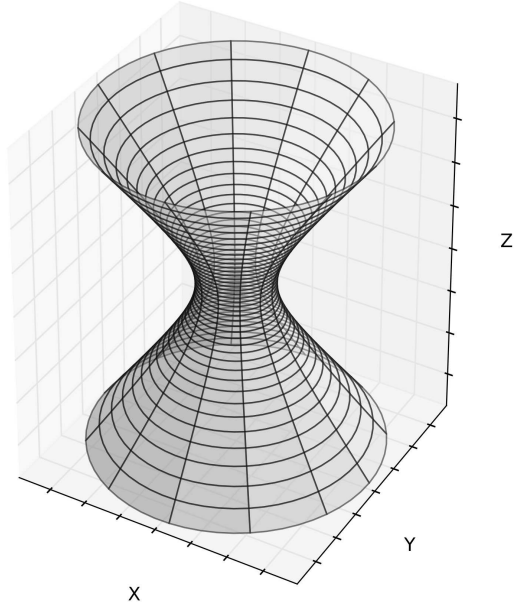}
\caption[hyperboloid]{\footnotesize The proposed constraint in Eq. (18) forces to the particles to move into an adequate geometry emulating sections of relativistic jets, specifically along of a hyperboloid as can be seen here for $Q>0$ and a small value of $P$ ($Q=0$ is for the conic case). Also, the maximum approach of the particles to the singularity is given when $r=\sqrt{Q}$  .}
\label{hyperboloid}
\end{center}
\end{figure}

\noindent As observations indicate, the jets propelled by black holes
consist of a collimated beam of relativistic particles \cite{Bridle}\cite%
{Belloni}. This suggests to constrain the movement of particles to an
adequate geometry that emulates sections of jets. In this direction, we add
a constriction to the Lagrangian (3) in such a way that tests particles are
forced to move along a hyperboloid (see Fig. \ref{hyperboloid}). As limit case, we first review the case
of movement in a cone. The revolution axis is $z$ and the opening cone angle 
$\theta $ should be small in order to emulate the mentioned beam collimation. Both cases are included in the surfaces generated by the relation $%
Ax^{2}+By^{2}-Pz^{2}-Q=0$. Here, we are assuming that $A$, $B$ and $P$ are
positive constants, while the conic and hyperboloid geometry are imposed by $%
Q=0$ and $Q>0$, respectively. Also, azimuthal symmetry is guaranteed by $%
A=B=1$. Then, without loss of generality we have as constraint:%
\begin{equation}
x^{2}+y^{2}-Pz^{2}-Q=0.  \tag{18}
\end{equation}%
Note that varying $P$ between $0$ and $1$ allows to vary $\theta $ from $0$
to $45%
{{}^\circ}%
$.

By mean of a Lagrange multiplier, we add the constraint (18) to the
Lagrangian (3), in the form

\begin{equation}
\mathcal{L} =\frac{1}{2}m\left[ -\gamma \dot{t}^{2}+\frac{\dot{r}^{2}}{%
\gamma }+r^{2}\left( \dot{\theta}^{2}+\sin ^{2}\theta \dot{\phi}^{2}\right) %
\right] +\frac{N}{2}\left( x^{2}+y^{2}-Pz^{2}-Q\right),  \tag{19}
\end{equation}%
where N is a Lagrange multiplier.

\noindent Lets first consider the case of a cone ($Q=0$). The idea is to follow
similar steps to those shown in section 2. Using this expression we find
that the Lagrangian (19) can be rewritten as%
\begin{equation}
\mathcal{L} =\frac{1}{2}m\left[ -\gamma \dot{t}^{2}+\frac{\dot{r}^{2}}{%
\gamma }+r^{2}\sin ^{2}\theta \dot{\phi}^{2}\right] +\frac{N}{2}[r^{2}(\sin
^{2}\theta -P\cos ^{2}\theta )].  \tag{20}
\end{equation}
Since in this expression we are assuming that the angle $\theta $ is
constant, we shall consider only variations of (20) with respect $t$, $r$, $%
\phi $ and $N$. From the Euler-Lagrange equation for $t$, we obtain $\frac{d%
}{d\tau }(m\gamma \dot{t})=0$ and therefore we find that

\begin{equation}
\dot{t}=\frac{a}{\gamma },  \tag{21}
\end{equation}%
where $a$ is a constant. After simplification, the variation of (21) with
respect to $r$ yields

\begin{equation}
\ddot{r}-\frac{\dot{r}^{2}\gamma ^{\prime }}{2\gamma }+\frac{\gamma \gamma
^{\prime }\dot{t}^{2}}{2}-\gamma r\left[ \sin ^{2}\theta \dot{\phi}^{2}+%
\frac{N}{m}(\sin ^{2}\theta -P\cos ^{2}\theta )\right] =0.  \tag{22}
\end{equation}%
The variation respect to $\phi $ leads to $\frac{d}{d\tau }(mr^{2}\sin
^{2}\theta \dot{\phi})=0$, and thus we obtain%
\begin{equation}
\dot{\phi}=\frac{b}{r^{2}\sin ^{2}\theta },  \tag{23}
\end{equation}%
where $b$ is another constant.

For the Lagrange multiplier $N$, it results 
\begin{equation}
r^{2}(\sin ^{2}\theta -P\cos ^{2}\theta )=0,  \tag{24}
\end{equation}%
which is consistent with the constraint $x^{2}+y^{2}-Pz^{2}=0$, obtained
from (18) when $Q=0$. Note that from this we get%
\begin{equation}
P=\tan ^{2}\theta .  \tag{25}
\end{equation}

By using (21), (23) and (24) the equation (22) becomes%
\begin{equation}
\ddot{r}-\frac{\dot{r}^{2}\gamma ^{\prime }}{2\gamma }+\frac{a^{2}\gamma
^{\prime }}{2\gamma }-\frac{b^{2}\gamma }{r^{3}\sin ^{2}\theta }=0.  \tag{26}
\end{equation}%
Note that if we make the association $h^{2}\rightarrow \frac{b^{2}}{\sin
^{2}\theta }$, (26) takes exactly the same form as (11). But now, the
Schwarzschild metric implies $-\gamma \dot{t}^{2}+\frac{\dot{r}^{2}}{\gamma }%
+r^{2}\sin ^{2}\theta \dot{\phi}^{2}=-1$, and by (21) and (23) this
expression can be written as%
\begin{equation}
\dot{r}^{2}=-\gamma +a^{2}-\frac{b^{2}\gamma }{r^{2}\sin ^{2}\theta }. 
\tag{27}
\end{equation}%
Insertion of (27) into (26) leads to

\begin{equation}
\ddot{r}+\frac{\gamma ^{\prime }}{2}+\frac{b^{2}\gamma ^{\prime }}{%
2r^{2}\sin ^{2}\theta }-\frac{b^{2}\gamma }{r^{3}\sin ^{2}\theta }=0. 
\tag{28}
\end{equation}

By following similar steps to the previous section, taking into account $%
\gamma =1-\frac{2M}{r}$ and $\gamma ^{\prime }=\frac{2M}{r^{2}}$, after a
straightforward computation we see that (27) yields to the integration
constant

\begin{equation}
\frac{\dot{r}^{2}}{2}-\frac{M}{r}+\frac{b^{2}}{2r^{2}\sin ^{2}\theta }-\frac{%
b^{2}M}{r^{3}\sin ^{2}\theta }=E.  \tag{29}
\end{equation}

The trajectories with angular momentum zero ($b=0$) correspond to radial
paths, and then (29) converts to%
\begin{equation}
\frac{\dot{r}^{2}}{2}-\frac{M}{r}=E.  \tag{30}
\end{equation}%
From (21) and (30) we have the components of the four velocity vector $%
u^{\alpha }$, $u^{\alpha }=\left( \frac{a}{\gamma },\pm \sqrt{-\gamma +a^{2}}%
,0,0\right) $(recall that $E=\frac{a^{2}-1}{2}$ and $\gamma =1-\frac{2M}{r}$%
), that clearly satisfies the normalization $u^{\alpha }u_{\alpha }=-1$. In
the component $u^{1}=\dot{r}$ the positive sign indicates particles going
outwards while the negative sign corresponds to free falling particles \cite%
{Hartle}.

From (30) we have that $\dot{r}=\pm \sqrt{\frac{2(Er+M)}{r}}$, with solution%
\begin{equation}
\sqrt{Er(Er+M)}-M\ln \left[ \sqrt{E(Er+M)}+E\sqrt{r}\right] =\pm E\sqrt{2E}%
\left( \tau -\tilde{\tau}\right) .  \tag{31}
\end{equation}

Since from (21) we have that $\frac{dr}{dt}=\frac{dr}{d\tau }\frac{d\tau }{dt%
}=\pm \left( 1-\frac{2M}{r}\right) \sqrt{\frac{2(Er+M)}{(2E+1)r}}$, in
Schwarzschild coordinates the solution is

\begin{equation}
\begin{array}{c}
\sqrt{2E+1}\left \{ \sqrt{Er(Er+M)}+(4E-1)M\ln \left[ \sqrt{E(Er+M)}+E\sqrt{r%
}\right] \right \} - \\ 
_{{}} \\ 
4ME\sqrt{2E}\tanh ^{-1}\left[ \sqrt{\frac{\left( 2E+1\right) r}{2(Er+M)}}%
\right] =\pm E\sqrt{2E}(t-\tilde{t}).%
\end{array}
\tag{32}
\end{equation}

The particular case where the velocity of the particle tends to zero in $%
r\rightarrow \infty $ corresponds to $E=0$ in (30), and correspondingly $%
\dot{r}=\sqrt{\frac{2M}{r}}$. Integration yields%
\begin{equation}
r=\sqrt[3]{2M\left[ \frac{3}{2}(\tau -\hat{\tau})\right] ^{2}}.  \tag{33}
\end{equation}

This result can be recasted in terms of Schwarzschild coordinates $(t,r)$,
since from (21) we have $\frac{dr}{dt}=\frac{dr}{d\tau }\frac{d\tau }{dt}=%
\sqrt{\frac{2M}{r}}\left( 1-\frac{2M}{r}\right) $, with solution

\begin{equation}
t=\hat{t}+4M\left[ \sqrt{\frac{r}{2M}}+\frac{1}{3}\left( \frac{r}{2M}\right)
^{3/2}-\tanh ^{-1}\sqrt{\frac{r}{2M}}\right] .  \tag{34}
\end{equation}

In (31-34) the quantities $\tilde{\tau}$, $\tilde{t}$, $\hat{\tau}$ and $\hat{t}$ are
integration constants.

These results correspond to the conic geometry [$Q=0$ in (18)]. This can be seen
as a limiting case for a model with a point source for the collimated beam
modelling a jet. It can be also interpreted as the limit case for a
hyperbolic constraint. Therefore, we focus the attention in the case $Q>0$
in (18). Thus, the more general Lagrangian now becomes%
\begin{equation}
\tciLaplace =\frac{1}{2}m\left[ -\gamma \dot{t}^{2}+\frac{\dot{r}^{2}}{%
\gamma }+r^{2}\left( \dot{\theta}^{2}+\sin ^{2}\theta \dot{\phi}^{2}\right) %
\right] +\frac{N}{2}[r^{2}(\sin ^{2}\theta -P\cos ^{2}\theta )-Q],  \tag{35}
\end{equation}%
where now $\dot{\theta}\neq 0$. The variations respect to $t$ and $\phi $
give again (21) and (23), while the variations respect to $r$ and $\theta $
results in%
\begin{equation}
\ddot{r}-\frac{\dot{r}^{2}\gamma ^{\prime }}{2\gamma }+\frac{\gamma \gamma
^{\prime }\dot{t}^{2}}{2}-\gamma r\left[ (\dot{\theta}^{2}+\sin ^{2}\theta 
\dot{\phi}^{2})+\frac{N}{m}(\sin ^{2}\theta -P\cos ^{2}\theta )\right] =0, 
\tag{36}
\end{equation}%
and%
\begin{equation}
\frac{d}{d\tau }(mr^{2}\dot{\theta})-mr^{2}\sin \theta \cos \theta \dot{\phi}%
^{2}-(1+P)Nr^{2}\sin \theta \cos \theta =0,  \tag{37}
\end{equation}%
respectively.

It is useful to express the constraint in (18) as%
\begin{equation}
r^{2}=\frac{Q}{\sin ^{2}\theta -P\cos ^{2}\theta }.  \tag{38}
\end{equation}

Multiplying Eq. (37) by $mr^{2}\dot{\theta}$ and inserting (23) and (38) in
its second and third term, we get rid of the dependence on $r$ and $\dot{\phi%
}$. The result is

\begin{equation}
mr^{2}\dot{\theta}\frac{d}{d\tau }(mr^{2}\dot{\theta})-m^{2}b^{2}\frac{\cos
\theta \dot{\theta}}{\sin ^{3}\theta }-mN(1+P)Q^{2}\frac{\sin \theta \cos
\theta \dot{\theta}}{[\sin ^{2}\theta -P\cos ^{2}\theta ]^{2}}=0.  \tag{39}
\end{equation}

This can be rewritten as

\begin{equation}
\frac{d}{d\tau }\left[ (mr^{2}\dot{\theta})^{2}+\frac{m^{2}b^{2}}{\sin
^{2}\theta }+\frac{mNQ^{2}}{\sin ^{2}\theta -P\cos ^{2}\theta }\right] =0. 
\tag{40}
\end{equation}

We use again (23) and (38) to eliminate $b$ and $Q$, and after integration
(40) gets converted to

\begin{equation}
(\dot{\theta}^{2}+\sin ^{2}\theta \dot{\phi}^{2})+\frac{N}{m}(\sin
^{2}\theta -P\cos ^{2}\theta )=\frac{l^{2}}{m^{2}r^{4}},  \tag{41}
\end{equation}%
where $l$ is constant. Now, this expression appears in square brackets in
Eq. (36), and then we have:

\begin{equation}
\ddot{r}-\frac{\dot{r}^{2}\gamma ^{\prime }}{2\gamma }+\frac{\gamma \gamma
^{\prime }\dot{t}^{2}}{2}-\frac{l^{2}\gamma }{m^{2}r^{3}}=0.  \tag{42}
\end{equation}

In a convenient way, we define $h=\frac{l}{m}$ and use [from (21)] $\dot{t}%
^{2}=\frac{a^{2}}{\gamma ^{2}}$ in the last relation, obtaining%
\begin{equation}
\ddot{r}-\frac{\dot{r}^{2}\gamma ^{\prime }}{2\gamma }+\frac{a^{2}\gamma
^{\prime }}{2\gamma }-\frac{h^{2}\gamma }{r^{3}}=0.  \tag{43}
\end{equation}

Again, we have the same relation as (11), although in this case
integration is less trivial. From the definition of solid angle, $\dot{\Omega%
}^{2}=$ $\dot{\theta}^{2}+\sin ^{2}\theta \dot{\phi}^{2}$, and then (41) can
be rewritten as

\begin{equation}
\dot{\Omega}^{2}=\frac{h^{2}}{r^{4}}-\frac{N}{m}(\sin ^{2}\theta -P\cos
^{2}\theta ).  \tag{44}
\end{equation}%
Also, the Schwarzschild metric implies the relation

\begin{equation}
-\gamma \dot{t}^{2}+\frac{\dot{r}^{2}}{\gamma }+r^{2}(\dot{\theta}^{2}+\sin
^{2}\theta \dot{\phi}^{2})=-\frac{a^{2}}{\gamma ^{2}}+\frac{\dot{r}^{2}}{%
\gamma }+r^{2}\dot{\Omega}^{2}=-1,  \tag{45}
\end{equation}%
where we have used $\dot{t}^{2}=\frac{a^{2}}{\gamma ^{2}}$. From the last
equality of the equation, isolating $\dot{r}^{2}$ and using (23) and (44),
results%
\begin{equation}
\dot{r}^{2}=a^{2}-\gamma \left[ 1+\frac{h^{2}}{r^{2}}-\frac{Nr^{2}}{m}(\sin
^{2}\theta -P\cos ^{2}\theta )\right] .  \tag{46}
\end{equation}
Then we substitute this in (43) to obtain

\begin{equation}
\ddot{r}+\frac{\gamma ^{\prime }}{2}\left[ 1+\frac{h^{2}}{r^{2}}-\frac{Nr^{2}%
}{m}(\sin ^{2}\theta -P\cos ^{2}\theta )\right] -\frac{h^{2}\gamma }{r^{3}}%
=0.  \tag{47}
\end{equation}

Using $\sin ^{2}\theta -P\cos ^{2}\theta =\frac{Q}{r^{2}}$[cf. Eq. (38)], $%
\gamma =1-\frac{2M}{r}$ and $\gamma ^{\prime }=\frac{2M}{r^{2}}$, (47) takes
the form

\begin{equation}
\ddot{r}+\frac{M}{r^{2}}\left( 1-\frac{NQ}{m}\right) =\frac{h^{2}}{r^{3}}(1-%
\frac{3M}{r}).  \tag{48}
\end{equation}
Multiplying by $\dot{r}$, this can be integrated to obtain the energy:

\begin{equation}
\frac{\dot{r}^{2}}{2}-\frac{M}{r}\left( 1-\frac{NQ}{m}\right) +\frac{h^{2}}{%
2r^{2}}-\frac{Mh^{2}}{r^{3}}=E.  \tag{49}
\end{equation}
Comparing with (28) -and considering $h=\frac{b}{\sin \theta }$-, we see that the
energies are very similar, where the difference is the single term $\frac{MNQ%
}{mr}$ in (49).\bigskip

\noindent \textbf{4. Final remarks.}\smallskip

\noindent In this work we have analysed the equations of motion corresponding
to conic and hyperbolic constraints in the trajectories of test particles in
a Schwarzschild geometry. We have obtained the expressions (23), (29), (41) and (49)  that can be
associated with (the conserved) energy, and angular momentum of the system.
In place of the geodesic equation, we have used directly the Lagrangian
approach, in order to add consistently the constraints. In the process, we
have verified that f restrictions in the orbits (such as angular
momentum zero) lead to the known solution to radial plunge orbits \cite%
{Hartle}.

As mentioned before, the constrained orbits analysed in this work can be
useful to model jets around black holes \cite{Smith}. With values of $P$
near to zero in Eq. (18), one can model a sufficiently collimated jet, while
relativistic corrections could be useful near the horizon of the black hole.
The hyperbolic constraint can be useful for jets originated near the
horizon, where the value $r_{\min }=\sqrt{Q}$ indicates the maximum approach
of the test particles to the singularity.

This work can be used to simplify approximations in models for jets (see for
instance \cite{Smith}-\cite{Fiziev} and references therein) around compact
astrophysical objects, and also could be related to other Lagrangian
approaches associated with black holes \cite{Hawking1}-\cite{Nieto}%
, but this is left for future research.\bigskip

\noindent \textbf{Acknowledgements}\smallskip

\noindent EAL would like to recognize finantial support by PROFAPI-UAS. JAN
would like to thank the CUCEI in Universidad de Guadalajara, for hospitality
during a stage of this work. Finally, ERL gratefully acknowledge a PhD
fellowship by CONACyT-Mexico.

\end{document}